# Thickness-dependent electron momentum relaxation times in iron films

K. L. Krewer,[1,2,a)] W. Zhang,[1,4] J. Arabski,[3] G. Schmerber,[3] E. Beaurepaire,[3] M. Bonn,[1] and D. Turchinovich[1,4]

[1]*Max Planck Institute for Polymer Research, 55128 Mainz, Germany*

[2]*Graduate School of Excellence Material Science in Mainz, 55128 Mainz, Germany*

[3]*Université de Strasbourg, CNRS, Institut de Physique et Chimie des Matériaux de Strasbourg, UMR 7504, 23 rue du Loess, F-67000 Strasbourg, France*

[4]*Fakultät für Physik, Universität Bielefeld, 33615 Bielefeld, Germany*

a) Corresponding author krewer@mpip-mainz.mpg.de

Terahertz time-domain conductivity measurements in 2 to 100 nm thick iron films resolve the femtosecond time delay between applied electric fields and resulting currents. This current response time decreases from 29 fs for thickest films to 7 fs for the thinnest films. The macroscopic response time is not strictly proportional to the conductivity. This excludes the existence of a single relaxation time universal for all conduction electrons. We must assume a distribution of microscopic momentum relaxation times. The macroscopic response time depends on average and variation of this distribution; the observed deviation between response time and conductivity scaling corresponds to the scaling of the variation. The variation of microscopic relaxation times depends on film thickness because electrons with different relaxation times are affected differently by the confinement since they have different mean free paths.

Conductivity in metals is typically described using a highly simplified model: a gas of identical electrons characterized by a single, universal relaxation time[1–4]. This contrasts with the complexity of the underlying process where all electronic states on the Fermi surface contribute to conduction[3,5], and relaxation times often vary strongly across the Fermi surface[6–10]. The few femtosecond time delay between an applied field and the resulting current - the current response time $\tau_C$ - reflects the macroscopic momentum relaxation of an ensemble of charges. Resolving this delay can therefore provide insights into the relaxation processes and connect microscopic scattering to macroscopic conduction.

This is particularly relevant for thin metal films[1,2,4,11–13] for which different electrons can be affected differently by confinement. Here, we determine $\tau_C$ as a function of thickness for thin metal films, and demonstrate that different types of electrons with different relaxation times are present in the films.

We performed substrate referenced transmission terahertz time-domain transmission spectroscopy[14,15] at room temperature (293 K) on iron films ranging from 2.2 to 100 nm thickness. The films were deposited on double-polished MgO (100) substrates and capped with ca. 12 nm of MgO. The films were grown by molecular beam epitaxy at room temperature, with subsequent annealing. The thicknesses $a$ were controlled in situ by quartz balance sensing and confirmed ex situ by small-angle x-ray diffraction (XRD) for selected samples (see supplementary material fig. S1). Roughnesses extracted from XRD average $0.9 \pm 0.1$ nm, without significant dependence on thickness. Reflection high-energy electron diffraction images indicate that this preparation method achieves single-crystalline films with bcc lattice structure (see fig. S2 and S3 in the supplementary material).

The terahertz radiation was generated and detected in 1 mm ZnTe crystals using 800 nm 40 fs pulses from an amplified Ti:Sapphire laser emitting 1000 pulses per second[14]. We alternated recording the terahertz transmission through the samples with the transmission through a bare reference substrate. We performed three rounds of measurements, each with a different combination of samples, alternatingly acquiring traces for sample and reference 10 to 30 times. We correct the terahertz transmission relative to the reference substrate for substrate thickness differences[16]. We then numerically solve the transfer matrices[17] for the corrected transmission data, using the thin conductive film approximation[18,19] to generate starting values. This approach allows reliably determining the phase $\varphi$ of the conductivity $\tilde{\sigma}$, even for films for which the phase acquired by the terahertz pulse during a direct transit is non-negligible (see Supplementary Material).

The current response time $\tau_C$ can be obtained from the measured phase $\varphi$ of the conductivity $\tilde{\sigma}$ at a specific frequency $f$, through:

$$\tau_C(f) = \tan\left(\varphi(\tilde{\sigma}(f))\right)/(2\pi f) \qquad (1)$$

If a universal relaxation time $\tau_u$ existed, the current response time $\tau_C$ would be constant and equal to the universal relaxation time $\tau_u$ of the Drude model. Previous measurements of the current response time in metals were limited to ca. 10 fs accuracy, due to uncertainties in the thickness of the reference substrate[20,21]. Our thickness correction technique allows determining the current response time with an error of ca. 1 fs[16] for most cases, allowing to compare response times between different samples. Fig.1 shows 3 exemplary phase-resolved conductivity spectra for a very thin (2.2 nm), an intermediate (10.3 nm), and a thick film (100 nm). The phase-resolved conductivity is plotted in terms of amplitude $\sigma$ and current response time $\tau_C$. The spectra are essentially flat, with the exception of the conductivity amplitude $\sigma$ of the 100 nm film, which slightly decreases with increasing frequency. Deviations from flat spectra are larger than the statistical errors and correlate between different samples; for example an increase in response time $\tau_C$ for high frequencies occurs for both the 2.2 and 10 nm films shown in fig. 1. These residuals only correlate between different samples measured in the same round; the residuals do not correlate with the same samples measured in a different round several months later with a different reference substrate (see supplementary fig. S4). Therefore we consider these residuals artefacts.

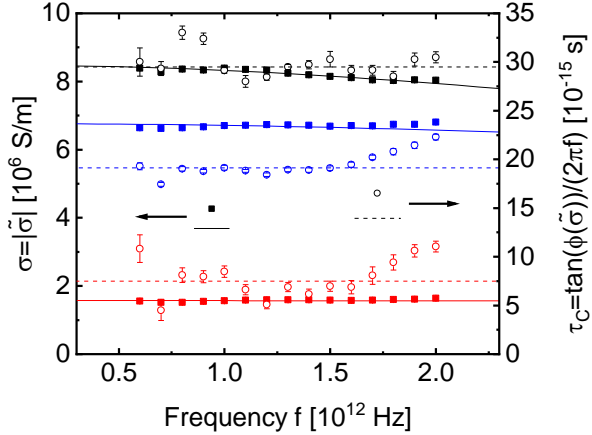

FIG. 1: The complex conductivities $\tilde{\sigma}$ extracted from time-domain spectroscopy, represented as amplitude $\sigma$ (full squares, left axis) and current response time $\tau_C$ (empty circles, right axis), which is the tangent of the phase over angular frequency. Conductivity spectra are displayed for a thick (100 nm, black), intermediate (10.3 nm, blue), and very thin (2 nm, red) sample. Error bars indicate the standard error of the mean, inferred from repeated measurements. Where invisible, the error bars are smaller than the markers. Lines denote effective Drude responses derived from spectral averages $\bar{\tau}_C$ (dashed lines, right axis) and $\bar{\sigma}_{DC}$ (full lines, left axis) of $\tau_C(f)$ and $\sigma(f) \cdot \sqrt{1 + (2\pi f \bar{\tau}_C)^2}$.

The amplitude spectra $\sigma(f)$ are flat due to the very low relaxation times $\tau$. The flat response time spectra $\tau_C(f)$ appear to be consistent with the hypothesis of a universal relaxation time $\tau_u$, as the hypothesis predicts the response time to be constant and equal to $\tau_u$, as a universal relaxation time results in a Drude type dispersion

$$\tilde{\sigma} = \sigma_{DC}/(1 - i2\pi f \tau_u). \qquad (2)$$

The universal relaxation time assumption also predicts that the DC conductivity $\sigma_{DC}$ is proportional to $\tau_u$. We find that the conductivity amplitudes $\sigma$ are lower for the thinner films, with the difference between 10 to 2.2 nm being much larger than that between 100 and 10 nm. The current response time $\tau_C$ behaves quite differently. It is also lower for thinner films, but the relative difference in response times between 100 and 10 nm is almost two times that of the conductivity amplitudes, contrary to the assumption of direct proportionality. The 2.2 nm thin film shows higher current response times than direct proportionality would predict from the conductivity amplitudes.

To compare the scaling of conductivity amplitude and response time, we extract the spectral average $\bar{\tau}_C$ of the current response times for each film. We then compute the DC-limit of the conductivity of a Drude type dispersion from each frequency step by multiplying the conductivity amplitudes $\sigma$ at each frequency $f$ with $\sqrt{1 + (2\pi f \bar{\tau}_C)^2}$. The resulting spectra for the DC parameter are then also averaged. The residual artefacts are the main source of error on both the response time $\bar{\tau}_C$ and DC-conductivity $\bar{\sigma}_{DC}$ parameters extracted from each spectrum. The residuals are taken into account by multiplying the variance of the weighted averages by the reduced sum of weighted residuals. Further, we use the instances where we have measurements of the same sample in different rounds to estimate the precision. These measurements deviate between 0.1 and 2.7 fs from another, i.e. slightly more than what we estimate from the residuals. We use the larger estimate where we have measurements from different rounds, and we add the average unexplained variance of (0.8 fs)[2] to the variance from the residuals where we do not have measurements from different rounds. Further, we add the variances caused by a possible 1% error in the substrate refractive index, by the uncertainty of the substrate thickness correction and the uncertainty on the film thickness $a$.

The thickness scaling of the extracted response time and DC-conductivity is shown in fig. 2. From thick towards thin films, the DC conductivity first hardly decreases, then jumps down from 10 to 8 nm and then keeps decreasing strongly. The current response time first decreases quickly, also jumps down from 10 to 8 nm and then levels out towards thinner films. So while both parameters decrease with decreasing thickness $a$, the detailed scaling is different.

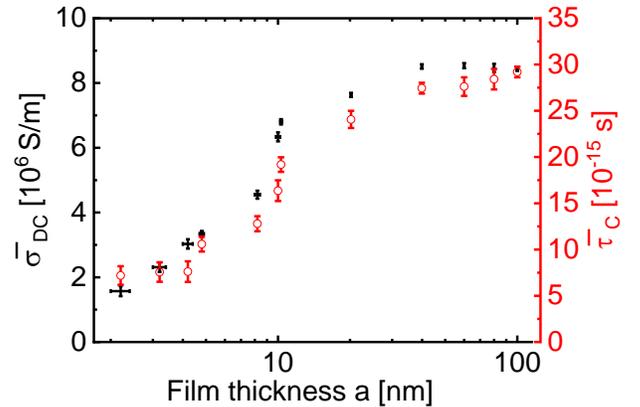

FIG. 2: DC-conductivities $\bar{\sigma}_{DC}$ (left axis, black) and current response times $\bar{\tau}_C$ (right axis, red) extracted from complex conductivity spectra of 12 iron films from 2.2 to 100 nm thickness. The conductivity decreases and the current response becomes faster for thinner metal films, as expected from increased surface scattering, but they are not directly proportional. All measurements were performed at 293 K.

None of the existing models for the thickness scaling of conductivity predict the response time[1,2,4,11–13]. Further, the jump between 8 and 10 nm does not fit with any model. We hence do not focus on trying to find a detailed microscopic model for the exact thickness scaling of these iron films. We rather focus on the peculiarity of seeing constant current response time spectra but no proportionality between conductivity and current response time. The spectrally constant response time is predicted by the universal relaxation time hypothesis; the deviations from proportionality contradict this hypothesis. To investigate the deviation from proportionality, we plot the quotient $Q = \bar{\tau}_C/\bar{\sigma}_{DC}$ of the response time and the DC-conductivity in fig. 3. We see that those values are not constant, but rather start out high, decrease down to 10 nm and increase again. We analyze how significant these deviations from proportionality are. We assume our data obey a normal distribution. Our error bars are our best estimates for the 68% confidence interval. The best fit for a constant quotient lies several interval widths above the 68 % confidence intervals for the intermediate films around 10 nm, and several times below those for 100 nm. The probability of obtaining data fitting worse to the hypothesis of a universal relaxation time is $10^{-9}$, equivalent to 6 standard deviations for a normal distribution. The deviations are significant.

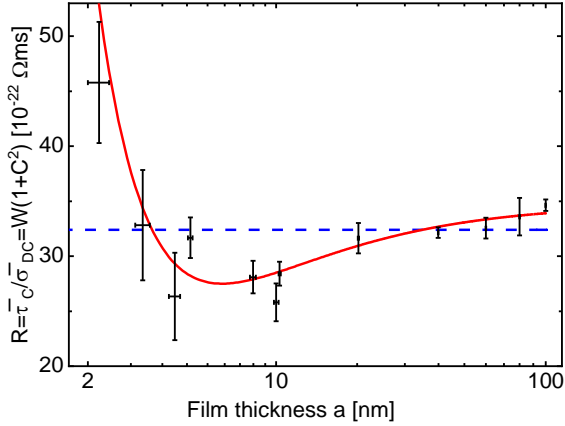

FIG. 3: Quotient $Q$ of the current response time $\tau_C$ by the DC-conductivity $\sigma_{DC}$ as a function of the film thickness. The blue dashed line is a constant fitted to the data, as predicted by the hypothesis of a universal relaxation time. The data lie outside the 99.999999% (6 Gaussian standard deviations) confidence interval for the universal relaxation time hypothesis. The red curve is a second-order polynomial fit in $1/a$, consistent with two competing processes altering the variation $C$ of the relaxation times. Data are within the 68 % (1 std) confidence interval of the red curve.

How can we explain now that the spectra of the response time are flat but it is not the universal relaxation time? Similar behavior has been observed by Kamal et al.[22] in a metal oxide film. The entire spectrum of the metal oxide can only be described by a distribution of scattering rates $\tau^{-1}$; but for the low frequency limit, a single effective current response time suffices[22]. Similarly, we can also understand our observation by considering a distribution of microscopic relaxation times $\tau$. We start by using the full description of the phase-resolved conductivity spectrum within the relaxation time approximation in the semi-classical Bloch-Boltzmann formalism. For a cubic crystal, this yields[3,6,23]:

$$\tilde{\sigma}(f) = \frac{1}{3}\frac{e^2}{8\pi^3}\sum_{b,s}\int_0^\infty \iint_{S(\mathcal{E})} -\frac{\partial g}{\partial \mathcal{E}} v_g(\vec{S},\mathcal{E},b,s) \frac{\tau(\vec{S},\mathcal{E},b,s)}{1-i2\pi f \tau(\vec{S},\mathcal{E},b,s)} dS d\mathcal{E}. \quad (3)$$

Here we sum over all bands $b$ and both spins projections $s$ and integrate over all iso-energy surfaces $S(\mathcal{E})$ in reciprocal space. The conductivity contribution for each point $\vec{S}$ on such an iso-energy surface is given by the group velocity $v_g$ and the relaxation time $\tau$, which both may vary between each point $\vec{S}$, energy $\mathcal{E}$, band $b$, and spin $s$.[6] The contribution is weighted by the energy derivative of the electron distribution function $g$; when the electron population conducts in a steady-state, this is the Fermi-distribution. The derivative of the Fermi distribution means that only electronic states close to the Fermi surface contribute. $g$ will deviate from the Fermi distribution when the energy of the photons $hf$ is larger than the typical thermal excitation $k_B T$. For room temperature (293 K) this implies frequencies $f$ larger than 6 THz. Since we stay below 6 THz, the observables we measure only depend on the same microscopic properties which govern steady-state transport at room temperature.

The key point is that already in the semiclassical theory, the relaxation time is not necessarily universal, but varies between the electronic states contributing to conduction. Since we want to assess the phase and frequency dependence of the conductivity, we convert the integral of eq. (3) into an integration over the relaxation times.

$$\tilde{\sigma}(f) = W^{-1} \int_0^\infty \frac{w(\tau)\tau}{1-i2\pi f\tau} d\tau \quad (4)$$

Basically all parameters $\vec{S}, \mathcal{E}, b$, and $s$ have been expressed as functions of $\tau$, allowing to express the integrand as a single function $w(\tau)$. The prefactor $W^{-1}$ is chosen such that $w(\tau)$ is normalised. This allows interpreting $w(\tau)$ as a probability distribution density. The distribution of microscopic relaxation times $w(\tau)$ gives the probability that a microscopic excitation of unit conductivity per relaxation time will relax in a time $\tau$. $W$ only depends on the electronic structure of the material and not on the relaxation times $\tau$. In the case of a universal relaxation time $\tau_u$, the distribution $w(\tau)$ reduces to a delta distribution $\delta(\tau - \tau_u)$. Inserting this in eq. (4) and looking at the DC-limit, we identify $W$ as the quantity $\rho_0 \tau_u$ calculated by Gall[3] by integrating computed group velocities over the Fermi surface. Within Drude's assumption of a universal relaxation time and photon energy independence of $\frac{\partial g}{\partial \mathcal{E}}$, $W$ is connected to Drude's plasma frequency $\omega_P$ via $W^{-1} = \varepsilon_0 \omega_P^2$.

We measure the low-frequency limit of $\tilde{\sigma}(f)$. Therefore we Taylor-expand eq. (4) for frequencies $f$ lower than $1/(2\pi\tau)$:

$$\tilde{\sigma}(f) = \frac{1}{W}\sum_{l=0}^\infty (i2\pi f)^l \langle \tau^{l+1} \rangle \quad (5)$$

$$= \frac{\sigma_{DC}}{1-i2\pi f \tau_C} + O((2\pi f \tau)^2) \quad (6)$$

Here $\langle \ \rangle$ denotes the average over the distribution of relaxation times $w(\tau)$. Equation (5) tells us that the conductivity is directly connected to the moments $\langle \tau^l \rangle$ of the relaxation time distribution $w(\tau)$. Theoretically, all moments could be inferred from the spectrum, and thereby the entire distribution. In practice, we can infer information about the first two moments: Equation (6) holds for

$$\sigma_{DC} = \langle \tau \rangle / W \quad \text{and} \quad \tau_C = \frac{\langle \tau^2 \rangle}{\langle \tau \rangle} = \langle \tau \rangle (1 + C^2). \quad (7)$$

The first moment, the mean relaxation time $\langle \tau \rangle$, gives us the average magnitude. The second centralised moment is the variance $V$. The standard deviation $\sqrt{V}$ measures the absolute width of the distribution. To decide how much impact the shape of the distribution has, we need to compare the standard deviation to the mean. This ratio $\sqrt{V}/\langle \tau \rangle$ is the coefficient of variation $C$. We hence can interpret the deviations of $Q = \bar{\tau}_C / \bar{\sigma}_{DC} = W(1 + C^2)$ from a constant value in fig. 3 in terms of a change in variation $C$ of the microscopic relaxation times.

Now we will show that the deviations of $Q$ from a constant value are not random, but depend systematically on film thickness $a$. To this end, we show that a simple polynomial function of the thickness $a$ describes the data reasonably well. This polynomial is $Q(a) = Q_\infty - b_1/a + b_2/a^2$, displayed as the red curve in fig. 3. The probability $p$ of obtaining data fitting worse to this polynomial is 0.21. The expectation value of this probability is 0.5. The $p$ value is within 0.34 (one Gaussian standard deviation) of its expectation value. This indicates a good fit. The fit parameters are $Q_\infty = 34.8 \cdot 10^{-22}$ Ωms, $b_1 = 9.2 \cdot 10^{-15}$ Ωs and $b_2 = 2.9 \cdot 10^{-6}$ Ωs/m. This description with a simple polynomial function shows that the thickness scaling of $Q$ is systematic.

Next, we describe two effects by which surface scattering may alter the variation $C$ of microscopic relaxation times and thereby qualitatively explain the thickness scaling of $Q$. Firstly, the anticorrelation of bulk and surface scattering may decrease $C$. Surface scattering predominantly affects electrons with a long expected free path, since those electrons are most likely to reach the surface instead of scattering in bulk. This anticorrelation starts cutting off the long relaxation time "tail" of the distribution of microscopic relaxation times, which decreases the variation. The thinner the film, the larger the role of surface scattering and the larger the reduction of the variation from bulk scattering by the anticorrelated surface scattering. We can interpret the $b_1$-term in the empirical polynomial as representing this narrowing of the relaxation time distribution by anticorrelated scattering mechanisms.

Secondly, the surface scattering will add positional ($z$) and directional ($v_z$) variation between microscopic relaxation times $\tau(\vec{S}, \mathcal{E}, b, s, z, v_z)$. An electron may scatter very soon from the surface when it is close to the surface and travels towards the surface. An electron that is far away from the surface or travelling parallel to the surface will hardly scatter from it. So the relaxation times due to surface scattering will vary strongly, and the directional variation will increase for increasingly thinner films. This variation will add to the intrinsic variation, increasing the total variation $C$. We can motivate the $b_2$ term by this effect. We should mention that contrary to the decrease of $Q$ from 100 to 8 nm, the strong rise of $Q$ for the thinnest films is much less significant due to large uncertainties. Further, competing explanations for this increase exist. Firstly, at thicknesses of a few nm, the roughness of almost 1 nm will cause systematic errors, as described by Namba[24]. Secondly, recent x-ray absorption measurements on a 1.5 nm film[25] suggest some changes to the electronic structure for such a thin film compared to bulk. This means for the thinnest films, the value of $W$ may change.

We now check whether our explanation of two mechanisms changing the variation of relaxation times is consistent with the observed conductivity scaling: When surface scattering dominates, the increase in directional variation will increase $Q$ with decreasing thickness. When bulk scattering dominates, the anticorrelation effect will decrease the variation with decreasing thickness. Therefore a minimum in variation $C$ and quotient $Q$ should occur when bulk and surface scattering contributions are about equal, which implies a conductivity of ca. half the bulk value. In our measurement, the conductivity drops to half the value of the thickest film between 5 and the 8 nm film thickness $a$, which coincides with the minimum the empirical $Q(a)$ scaling. So we have found a qualitative explanation for how surface scattering can cause the scaling of the variation $C$ and therefore the observed scaling of $Q$, and this explanation is consistent with the scaling of the conductivity.

Last but not least, we check the consistency of the variation scaling explanation by comparing $Q_\infty$ to values of $W$ calculated from electronic band structure. The anticorrelation between bulk and surface scattering can only take effect when a large variation $C_\infty$ exist in bulk, which implies that $Q_\infty$ must be larger than $W$. Cazzaniga et al.[26] use density functional theory with local spin density approximation to calculate a dc-conductivity $\sigma_{DC}$ of $155 \cdot 10^6$ S/m for an assumed universal relaxation time $\tau_u$ of 143 fs, from which we can calculate a value for $W = \tau_u/\sigma_{DC}$ of $9.2 \cdot 10^{-22}$ Ωms. For nickel, a cubic ferromagnet like iron, Gall[3] reports a similar calculated value of $10.0 \cdot 10^{-22}$ Ωms. Cazzaniga's value for $W$ is 3.8 times smaller than the $34.8 \cdot 10^{-22}$ Ωms we observe for $Q_\infty = W(1 + C_\infty^2)$. This translates to variation $C_\infty$ in bulk of 1.7; that means the stardard deviation of the relaxation time distribution is 1.7 times larger than the mean. This variation would certainly be large enough for narrowing of the relaxation time distribution by anticorrelated surface scattering to occur.

We illustrate the impact of the variation of microscopic relaxation times on the current response time by using Cazzaniga's value for $W$ to estimate the mean relaxation time $\langle\tau\rangle$ from the conductivity of the 100 nm film. We estimate 8 fs, while the observed current response time $\bar{\tau}_C$ is 29 fs.

In summary, we have resolved the phase of the THz conductivity of iron films with enough precision to demonstrate decreasing current response times $\bar{\tau}_C$ with decreasing thickness. We further could resolve significant deviations between the scaling of the DC-conductivity and the response time. This can happen when no universal relaxation time $\tau_u$ exists. At this point, we need to distinguish between the observable macroscopic response time $\tau_C$, the various microscopic relaxation times $\tau$, and the mean relaxation time $\langle\tau\rangle$ parameterizing DC-conduction. The conductivity spectrum can be fully described by the distribution $w(\tau)$ of microscopic relaxation times. For the low-frequency limit, the observable DC-conductivity depends on the mean $\langle\tau\rangle$, the response time on mean $\langle\tau\rangle$ and variation $C$ of the distribution of relaxation times. We explain the thickness scaling of the quotient $Q = \tau_C/\sigma_{DC}$ by surface scattering changing the shape of the distribution of relaxation times. The relaxation time distribution picture allows predicting the response time $\tau_C$ for other metallic systems: Mott[7] explained the temperature-independent conduction in Constantan and Manganin alloys by a process that changes the distribution, but not the average relaxation time. Therefore, the response time $\tau_C$ should increase with temperature in these alloys.

Supplementary material:

See supplementary material for iron film characterization, terahertz data treatment and spectral averaging/residual spectra.

Acknowledgements:
We are grateful to Eduard Unger for automatizing the measurements and to Zoltan Mics and Ivan Ivanov for building the high precision set-up used. K.K. acknowledges the support from MAINZ - Graduate School of Excellence Material Science in Mainz. D.T. acknowledges the project "Nonequilibrium dynamics in solids probed by terahertz fields" funded by the Deutsche Forschungsgemeinschaft (DFG, German Research Foundation) - Projektnummer 278162697 - SFB 1242.

Credit Line:

# Supplementary material for thickness-dependent electron momentum relaxation times in iron films

Keno L. Krewer,[1,2,*] Wentao Zhang,[1,4,] Jacek Arabski,[3] Guy Schmerber,[3] Eric Beaurepaire,[3] Mischa Bonn,[1] and Dmitry Turchinovich[1,4,]

[1]*Max Planck Institute for Polymer Research, 55128 Mainz, Germany*
[2]*Graduate School of Excellence Material Science in Mainz, 55128 Mainz, Germany*
[3]*Université de Strasbourg, CNRS, Institut de Physique et Chimie des Matériaux de Strasbourg, UMR 7504, 23 rue du Loess, F-67000 Strasbourg, France*
[4]*Fakultät für Physik, Universität Bielefeld, 33615 Bielefeld, Germany*

(2019-01-23)

### Sample preparation:

The iron films were deposited on double polished MgO (100) substrates and capped with ca. 12 nm of MgO. The molecular beam epitaxy was performed at room temperature with subsequent annealing at 600 K. The deposition rate was 0.05 nm/min at standard $10^{-10}$ Torr ($10^{-8}$ Pa) pressure. The thicknesses were controlled "in situ" by quartz balance sensing and confirmed "ex situ" by small-angle x-ray diffraction (XRD) for selected samples by means of a Rigaku SmartLab X-ray diffractometer equipped with a monochromatic source (Ge(220)×2) delivering a Cu Kα1 incident beam (45 kV, 200 mA, λ = 0.154056 nm), see Figure S1). Roughnesses extracted from XRD average 0.9±0.1 nm, without significant dependence on thickness. X-ray diffraction patterns (fig. S.2) and reflection high energy electron diffraction images (fig. S.3) indicate that this preparation method achieves single-crystalline films with bcc lattice structure.

### Extraction of THz conductivity spectra from field transmission spectra:

We note that this is the first time, that the complex conductivity has been recovered for "non-thin" metal films, that is films which the complex phase acquired by the terahertz field during a direct transit is non-negligible. We have expanded the substrate thickness correction determination we presented in a previous work[1] accordingly, but this yields deviations smaller than the statistical error. However, the full transfer matrix approach for recovering the complex conductivity is crucial for all films thicker than 20 nm, because the thicker the film, the smaller the current response time would appear in the conventionally used thin film approximation.

### Extraction of current response time and DC-conductivity from THz conductivity spectra

We extract the current response times and DC conductivities of the effective Drude model (eq. S1).

$$\tilde{\sigma}(f) = \frac{\sigma_{DC}}{1 - i2\pi f \tau_C}. \text{(S1)}$$

The frequency range between 0.6 and 2.0 THz has the best phase resolution. We solve the complex conductivity at each frequency for response time and DC-conductivity in this region. The resulting values are similar for all frequencies, but the values vary more than their standard errors from repeated measurements predict. We note that the discrepancies correlate between samples measured in the same measurement round, but not between the same sample measured in two separate rounds (see fig. S4.). Since we changed reference substrates between rounds, we suspect contamination/roughness of the reference substrates as main cause for these small discrepancies. We extract the current response times and DC-conductivities by averaging over the 0.6 to 2.0 THz frequency range. Experimental noise, discrepancies between frequencies and discrepancies between measurements a year apart at slightly different positions have been factored into the parameter and error estimates.

### Accuracy estimation for response time and DC-conductivity

We focus the discussion on the response time, the conductivity was handled in a similar manner. We estimate the impact of the artefacts in two ways: First by taking into account the reduced sum of residuals to estimate the statistical error on the average. Second we use the differences between the spectral averages of the same samples measured in different series. The statistical error on the measurements lead to errors between 0.2 and 1.2 fs. The differences between measurements of the same sample are between 0 and 2.6 fs, their standard deviation is ca. 1.7 fs. This is somewhat larger than what to expect from the statistical error, even when factoring in the additional error from the substrate thickness correction. Therefore, for the samples with more than one measurement, the measurements are averaged and the differences taken into account to determine the error. The average additional error deduced from comparing measurements of the same sample is 0.8 fs, and this average value is added into the error estimate for the samples with only one measurement. Additionally systematic errors from inaccuracy of the substrate index are factored in, for thick films also any uncertainties on the film thickness. The biggest error contributions are the errors inferred from the residuals and the differences between different rounds of measurements. Only in case of the 3 nm sample, the error from the substrate thickness difference surpasses them.

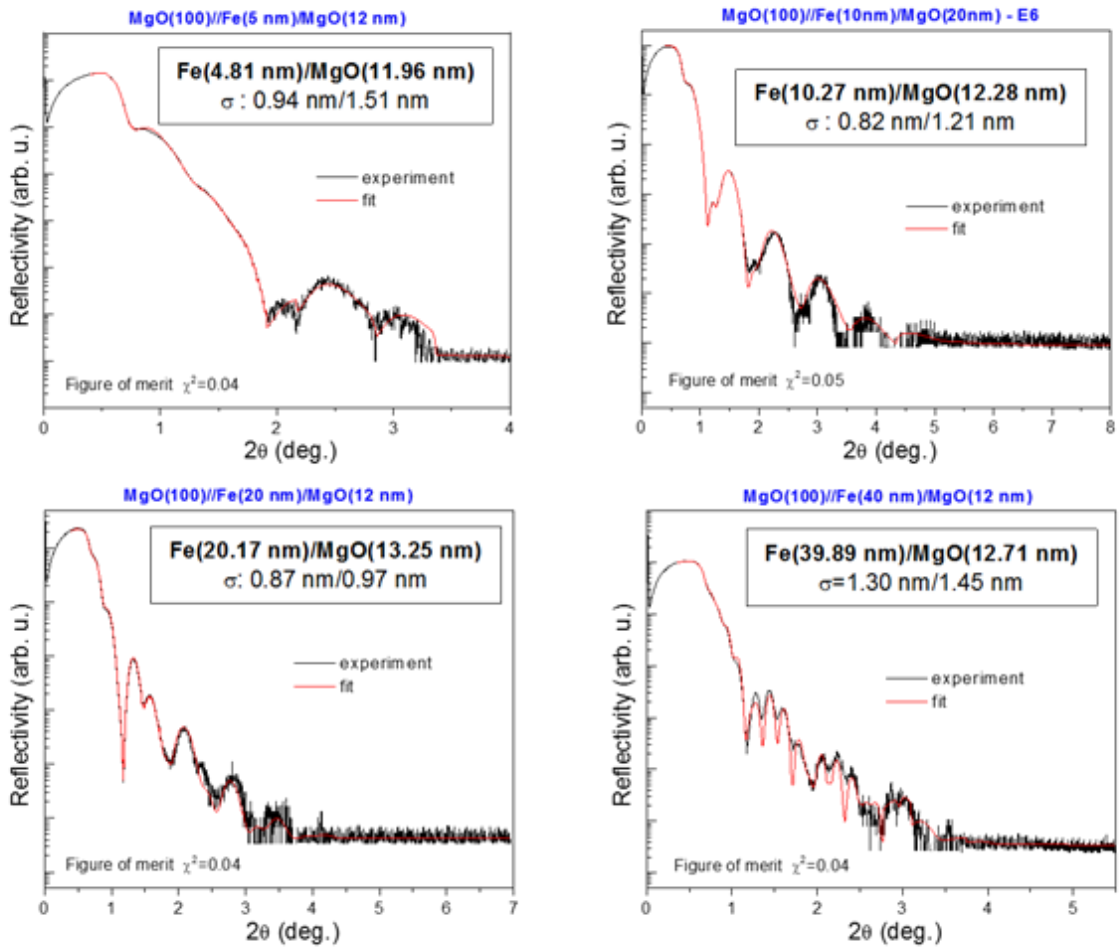

**Figure S1.** X-Ray reflectivity measurement of selected Fe thin films. The recovered thicknesses are given in parenthesis behind the layer materials. Below them, the respective roughnesses $\sigma$ are displayed. Experimental data points are shown in black, while the fit in red with a maximum figure of merit $\chi 2 = 0.05$.

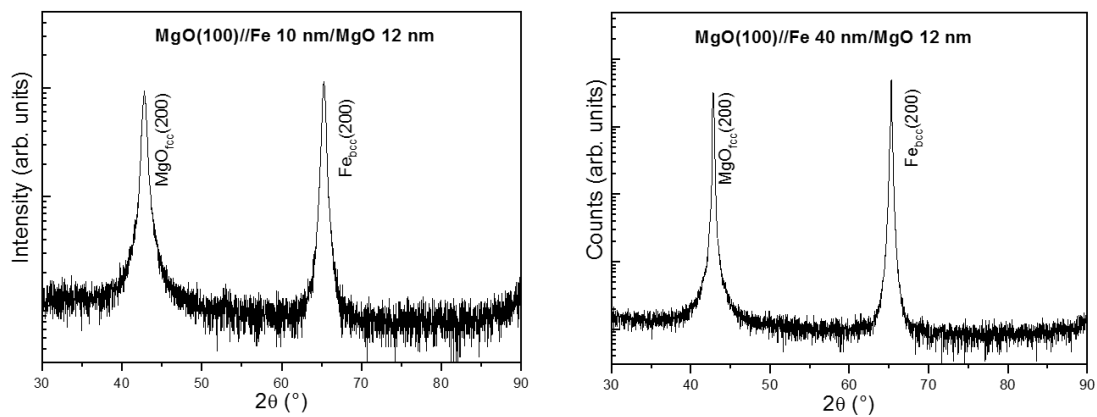

**Figure S2.** X-Ray diffraction pattern of selected Fe thin layers, Fe crystallizes well in the bcc structure.

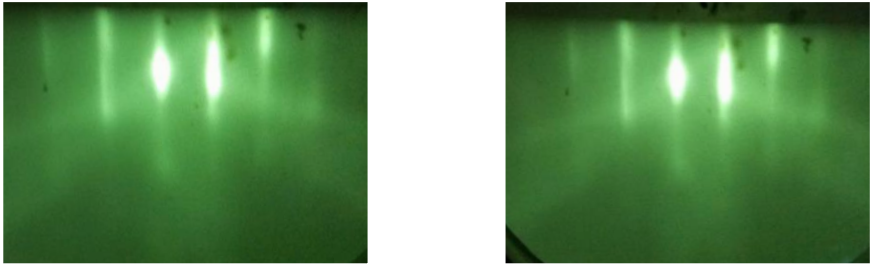

**Figure S3:** RHEED images of Fe on MgO/MgO epitaxial uptake.

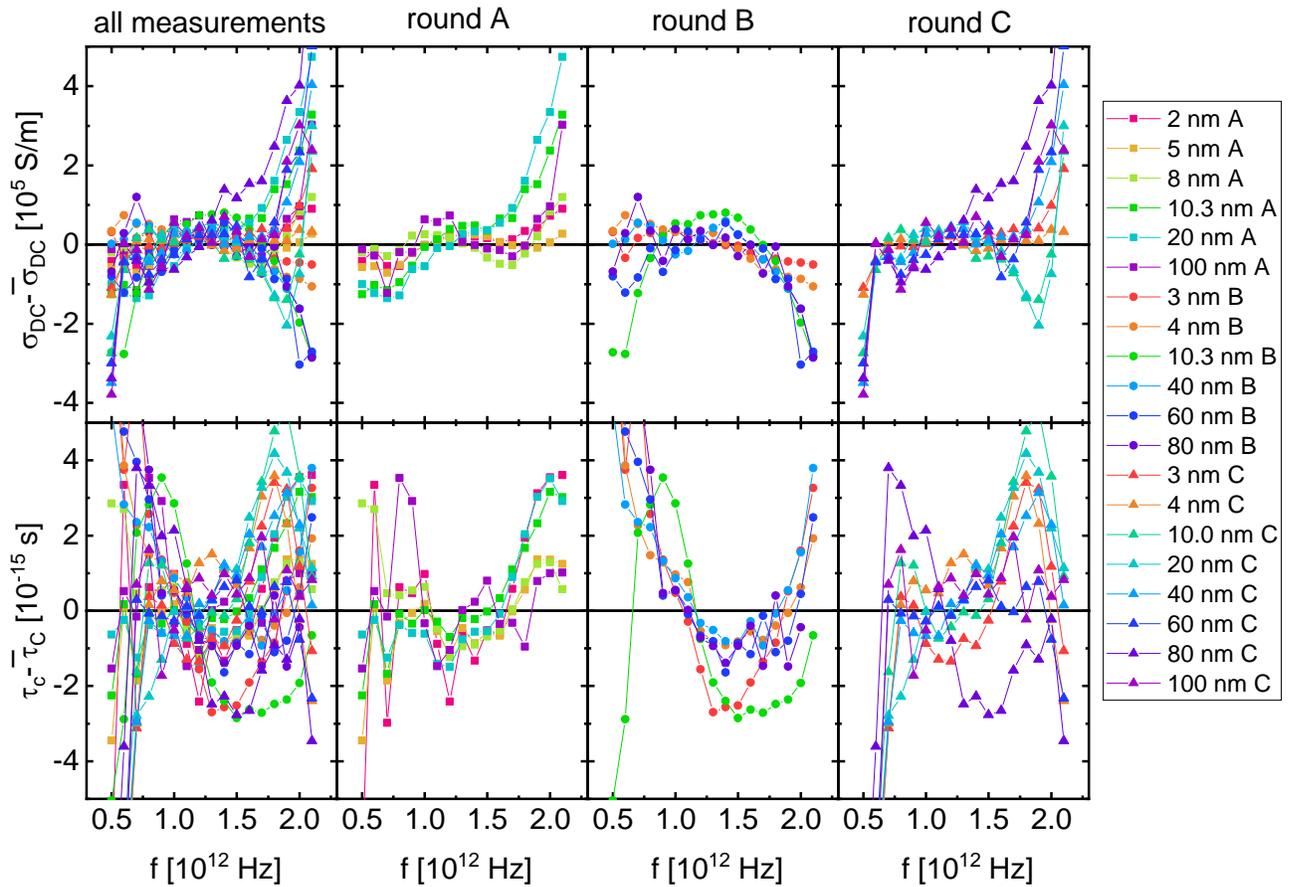

**Figure S4**: Spectra of the residuals of the spectral averages. Upper panels show the deviations between the DC conductivities $\sigma_{DC}(f) = \sigma(f) \cdot \sqrt{(1+(2\pi f \bar{\tau}_C)^2)}$ and their spectral averages $\bar{\sigma}_{DC}$. Lower panels show deviations of the current response times from their spectral averages. The first column shows the residuals of all spectra combined, columns 2 to 4 show each the spectra from a single measurement round. We can see clear trends among individual rounds, but no correlation between the rounds and none between the same samples or samples of similar thickness (similar thicknesses are indicated by similar color). The residuals are hence artifacts specific to the round of measurements.